\documentclass{jpsj-suppl}
\usepackage{txfonts}
\usepackage{bbm}

\renewcommand{\vec}[1]{\boldsymbol{#1}}

\newcommand{\be}{\begin{equation}}
\newcommand{\ee}{\end{equation}}
\newcommand{\bea}{\begin{eqnarray}}
\newcommand{\eea}{\end{eqnarray}}
\newcommand{\panda}{$\overline{\mbox P}$ANDA~}

\title{Production and Interactions of Hyperons and Hypernuclei}

\author{Horst \textsc{Lenske}$^{1,2}$, Xu \textsc{Cao}$^{3,4}$, Madhumita \textsc{Dhar}$^{1}$, Theodoros \textsc{Gaitanos}$^{1,5}$, and Radhey \textsc{Shyam}$^{6}$}

\inst{$^{1}$Institut f\"ur Theoretische Physik, JLU Giessen, Heinrich-Buff-Ring 16, D-35392, Giessen, Germany \\
$^2$GSI Darmstadt, D-64291 Darmstadt, Germany\\
$^3$Institute of Modern Physics, Chinese Academy of Sciences, Lanzhou 730000, China\\
$^4$State Key Laboratory of Theoretical Physics, Institute of Theoretical Physics, Chinese
Academy of Sciences, Beijing 100190, China\\
$^{5}$Aristotle University of Thessaloniki, Thessaloniki, Greece\\
$^6$Saha Institute of Nuclear Physics, Kolkata 700064,  India.}

\email{horst{\_}lenske@physik.uni-giessen.de}

\recdate{today}

\abst{The production of strangeness on the nucleon and hyperon and hypernuclear production in heavy ion collisions at relativistic energies and in antiproton annihilation on nuclei is discussed. The reaction process is described by transport theory with focus on $S=-2$ channels and a comparison of different model interactions. The interactions of hyperons in nuclear matter is investigated in a novel SU(3) approach. An outlook to the $S=-3$ sector and $\Omega^-$ physics is given.}

\kword{Strangeness, hyperons, hypernuclei, heavy ion collisions, transport theory, in-medium interactions }

\begin{document}
\maketitle

\section{Introduction}\label{sec:intro}

Hyperons and hypernuclei are short-lived objects and as such do not exist as stable particles. Their production always requires special efforts, as reviewed e.g. in \cite{Hash:2006}. High energy hadronic reactions provide a broad spectrum of final fragments, thus being a very suitable tool for investigations of strangeness in the nuclear medium. Below, we consider as an example strangeness production in elementary hadronic $(\pi,K)$ reactions on the nucleon. The production mechanism is described in the Giessen resonance model. The target nucleon is excited into a nucleon resonance $N^*$ which is located above the strangeness production threshold and decays to a hyperon $Y$ and an antikaon $K$. This resonance mechanism is also an important scenario in hyperon production in heavy ion collisions. The dense and hot environment in a heavy ion collision at energies of a few GeV per nucleon allows additional production paths by secondary rescattering processes of hadrons already produced in earlier stages of the reaction. The short-lived compressed and heated medium created in the interaction zone will expand and cool and finally the baryons will form fragments. This rather complex evolution is described in the initial phases by transport theory, supplemented by a statistical multi-fragmentation approach for the final step of fragment formation.

The production process is intimately connected to the interaction of hyperons with background medium by setting the conditions for the existence of hypernuclear bound states, as in the case of the $\Lambda$ hyperon, or the non-existence of a particle-stable nuclear system like for the $\Sigma$ baryons. Nucleons and hyperons together are forming the ground state baryon $SU(3)$ flavor octet. SU(3) symmetry, however is broken as becomes clear by the spread of about $400$~MeV among the octet baryon masses. Of course, broken SU(3) symmetry is taken onto account theoretically, at least on the level of masses and the corresponding production thresholds. In this work, we study octet baryon interactions in free space and cold infinite matter and in the context of heavy ion collisions which is a highly dynamical environment.

\section{Strangeness production in elementary reactions}\label{sec:Nucleon}

\begin{figure}[tbh]
\begin{center}
\includegraphics[width=14cm]{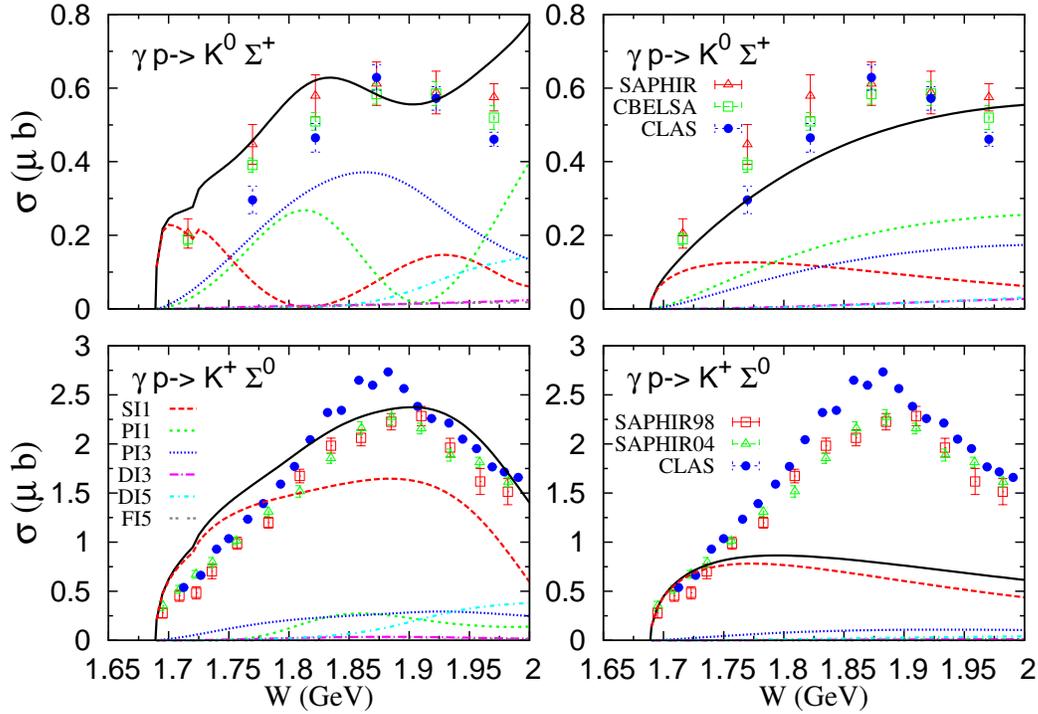}
\caption{Total cross sections for kaon production on the nucleon. Results of the Giessen model are compared to CLAS, CBELSA, and SAPHIR data \protect\cite{Xu:2013}.}
\label{fig:Fig1}
\end{center}
\end{figure}

Strangeness production on the nucleon plays a key role for our understanding of baryon structure and elementary reactions with hadrons. In addition, such production reactions are also an appropriate tool to identify excited states $N^*$ of the nucleon which decay preferentially into hyperon-kaon channels, thus adding to solve the problem of missing resonances. In \cite{Xu:2013} we have analysed the latest CLAS-, CBELSA-data sets and re-analysed the earlier SAPHIR-data on photoproduction of kaons on the nucleon. The Giessen model was used, describing meson production on the nucleon by a coupled channels K-matrix approach including meson production by photo-production and pion-induced reactions as initial channels. The Giessen model is obeying the elementary symmetries of hadron physics and conserving unitarity. Meson production proceeds through s-channel resonances and t- and u-channels re-scattering processes. Results for total cross sections are shown in Fig.\ref{fig:Fig1}. Up to a total center-of-mass energy of about $\sqrt{s}=2$~GeV the data are well described.

\begin{figure}[tbh]
\begin{center}
\includegraphics[width=12cm]{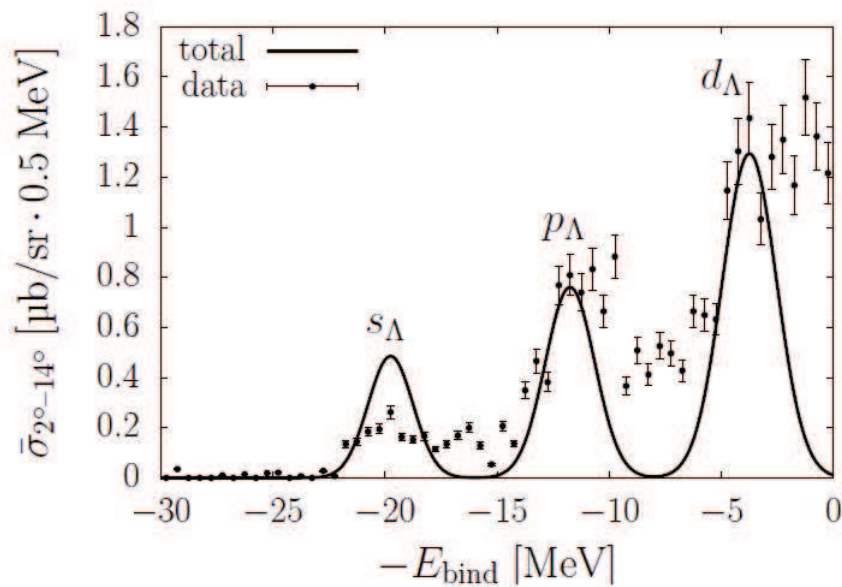}
\caption{Total cross section for kaon production in a $(\pi^+,K^+)$ reaction on $^{41}V$ at $P_{lab}=1.05$~GeV/c. Results of the Giessen resonance model \protect\cite{Bender:2009} are compared to KEK data \protect\cite{Hotchi:2001}. As indicated, s-, p-, and d-orbits are resolved.}
\label{fig:Fig2}
\end{center}
\end{figure}

Meson production through resonance excitation is one of the essential ingredients of the Giessen model. This mechanism has also been used in pion-induced kaon production on nuclei. A fully quantum mechanical approach was developed a few years ago in \cite{Bender:2009}. In the resonance model, including $\frac{1}{2}^\pm,\frac{3}{2}^\pm$ nucleon excitations, the production vertices are described by pseudo-scalar and pseudo-vector interaction Lagrangians \cite{Bender:2009}. The initial and final meson-baryon scattering channels were treated by solving the corresponding Klein-Gordon wave equations in eikonal approximation. Typical results are shown in Fig.(\ref{fig:Fig2}) where $^{51}V(\pi^+,K^+)^{51}_\Lambda V$ production cross sections are shown and compared to the KEK-data of Hotchi \textit{et al.} \cite{Hotchi:2001}. Nuclear wave functions are obtained in the relativistic DDRH mean-field model with density dependent coupling constants as discussed in \cite{Keil:2002,Lenske:2004}. The major contributions were found to be through excitation of three nucleon resonances, $N^*$(1650)[$\frac{1}{2}^-$], $N^*$(1710)[$\frac{1}{2}^+$], and
$N^*$(1720)[$\frac{3}{2}^+$], which have dominant branching ratios for
the decay to the $ K^+ \Lambda$ channel. The same approach was applied before to hypernuclear production in proton- and photon-induced reactions \cite{Shyam:2004,Shyam:2006}. These results as well as similar investigations found in the literature, e.g. \cite{Hiyama:2000}, illustrate that the basic mechanism of strangeness production in reactions on the free nucleon and on nuclei is understood essentially on a quantitative level.

\section{Interactions in the baryon octet}

It's almost trivial to state that the quantitative understanding of hypernuclei requires an accurate knowledge of hyperon-hyperon $(YY)$ and hyperon-nucleon $(YN)$ interactions. However, the solution of this basic task is still far from being solved satisfactorily to the degree of accuracy required for a hypernuclear theory of predictive power. The progress made in the last decade or so for $S=-1$ systems is only part of the full story since this involves mainly single-$\Lambda$ hypernuclei in addition to the few data points from old $p\Lambda$ and $p\Sigma$ experiments, e.g. in \cite{Ale:1968,Sec:1968,Eis:1971}. The latter are essential input for the approaches developed over the years by several groups. While the Nijmegen \cite{Nijmegen,Yama:2014} and the Juelich group \cite{Juelich}, respectively, are using a baryon-meson approach, the Kyoto-Niigata group \cite{Niigata} favors a quark-meson picture, finally ending also in meson-exchange interactions. However, none of the existing parameter sets is in any sense constrained with respect to interactions in the $|S|\geq 2$ channels. As a consequence, the implementation of SU(3) flavor symmetry is incomplete. In view of this situation, we found it necessary to reconsider interactions in the lowest SU(3) baryon octet with special emphasis on in-medium interactions. As a long-term program the approach will allow to test interactions more directly under the dynamical conditions of strangeness and hypernuclear production reactions with leptonic, hadronic, and nuclear probes, up to heavy ion reactions at relativistic energies. Hyperon and hypernuclear data from reactions are providing a wealth of information which has not yet been taken advantage of in full depth.

SU(3) octet physics is based on treating the eight baryons on equal footing. Although SU(3) flavor symmetry is broken by about 20\% on the mass scale, it is still meaningful to exploit the relations among coupling constants, thus reducing the number of free parameters considerably. The eight $J^P={\textstyle\frac{1}{2}}^+$ baryons are collected into a traceless matrix $B$, which is given by a superposition of the eight Gell-Mann matrices $\lambda_i$, leading to the familiar form
\begin{equation}
  B = \left( \begin{array}{ccc}
      {\displaystyle\frac{\Sigma^{0}}{\sqrt{2}}+\frac{\Lambda}{\sqrt{6}}}
               &  \Sigma^{+}  &  p  \\[2mm]
      \Sigma^{-} & {\displaystyle-\frac{\Sigma^{0}}{\sqrt{2}}
                   +\frac{\Lambda}{\sqrt{6}}}  &  n \\[2mm]
      -\Xi^{-} & \Xi^{0} &  {\displaystyle-\frac{2\Lambda}{\sqrt{6}}}
             \end{array} \right),
\end{equation}
and which is invariant under SU(3) transformations. Similarly, the pseudo-scalar($Ps$), vector ($V$), and, last but not least, the scalar ($S$)  meson nonets are expressed correspondingly. Taking the pseudo-scalar mesons with $J^P=0^-$ as an example we obtain the octet matrix $P_{8}$
\begin{equation}
   P_{8} = \left( \begin{array}{ccc}
      {\displaystyle\frac{\pi^{0}}{\sqrt{2}}+\frac{\eta_{8}}{\sqrt{6}}}
             & \pi^{+}  &  K^{+}  \\[2mm]
      \pi^{-} & {\displaystyle-\frac{\pi^{0}}{\sqrt{2}}
         +\frac{\eta_{8}}{\sqrt{6}}}  &   K^{0} \\[2mm]
      K^{-}  &  \overline{{K}^{0}}
             &  {\displaystyle-\frac{2\eta_{8}}{\sqrt{6}}}
             \end{array} \right).
\end{equation}
which, for the full nonet, has to be completed by the singlet matrix $P_0$, given by the unit matrix multiplied by $\eta_0/\sqrt{3}$. We define the SU(3)-invariant combinations
\begin{equation}
  \left[\overline{B}BP_8\right]_{F}= {\rm Tr}\left(\left[\overline{B},B\right]_{-}P_{8}\right)\quad , \quad
  \left[\overline{B}BP_8\right]_{D}
= {\rm Tr}\left(\left[\overline{B},B\right]_{+}P_{8}\right)\quad ,\quad \left[\overline{B}BP_0\right]_{S} = {\rm Tr}(\overline{B}B){\rm Tr}(P_{0})
\end{equation}
where $F$ and $D$  correspond to anti-symmetric and symmetric combinations of field operators, as indicated by the commutators and anti-commutators, $[A,B]_\mp$, respectively. With these relations we obtain the interaction Lagrangian
\begin{equation}
   {\cal L}_{I} = -f_{8}\sqrt{2}\left\{
     \alpha\left[\overline{B}BP_8\right]_{F}+
     (1-\alpha)\left[\overline{B}BP_8\right]_{D}\right\}\, - \,
     f_{0}{\textstyle\sqrt{\frac{1}{3}}}
     \left[\overline{B}BP_0\right]_{S},             \label{LIsu3}
\end{equation}
where $\alpha=F/(F+D)$ is a ratio of generic SU(3) coupling constants. The square-root factors are introduced following a commonly used convention. We introduce the isospin doublets
\begin{equation}
  N=\left(\begin{array}{c} p \\ n \end{array} \right), \ \ \
  \Xi=\left(\begin{array}{c} \Xi^{0} \\ \Xi^{-} \end{array} \right), \ \ \
  K=\left(\begin{array}{c} K^{+} \\ K^{0} \end{array} \right),
  \ \ \   K_{c}=\left(\begin{array}{c} \overline{K^{0}} \\
               -K^{-} \end{array} \right) \quad .        \label{eq:doublets}
\end{equation}
The $\Sigma$ hyperon and the pion isovector-triplets are chosen as
\begin{equation}
  \vec{\Sigma}\!\cdot\!\vec{\pi} = \Sigma^{+}\pi^{-}+\Sigma^{0}\pi^{0}+\Sigma^{-}\pi^{+}.
\end{equation}
also serving to fix phases. We define the pseudo-vector Dirac-vertex operator $\Gamma=\gamma_{5}\gamma_{\mu}\partial^{\mu}$. By evaluating the $F$- and $D$-type couplings, Eq.~(\ref{LIsu3}), we obtain the pseudo-scalar octet-meson interaction Lagrangian
\begin{eqnarray}
   m_{\pi}{\cal L}_{8} &\sim&
  -f_{N\!\pi}(\overline{N}\Gamma\vec{\tau}N)\!\cdot\!\vec{\pi}
  +if_{\Sigma\Sigma\pi}(\overline{\vec{\Sigma}}\!\times\!\Gamma\vec{\Sigma})
      \!\cdot\!\vec{\pi}
  -f_{\Lambda\Sigma\pi}(\overline{\Lambda}\Gamma\vec{\Sigma}+
      \overline{\vec{\Sigma}}\Gamma\Lambda)\!\cdot\!\vec{\pi}
  -f_{\Xi\Xi\pi}(\overline{\Xi}\Gamma\vec{\tau}\Xi)\!\cdot\!\vec{\pi}
            \nonumber\\
 &&-f_{\Lambda N\!K}\left[(\overline{N}\Gamma K)\Lambda
         +\overline{\Lambda}\Gamma(\overline{K}N)\right]
   -f_{\Xi\Lambda K}\left[(\overline{\Xi}\Gamma K_{c})\Lambda
         +\overline{\Lambda}\Gamma (\overline{K_{c}}\Xi)\right] \nonumber\\
 &&-f_{\Sigma N\!K}\left[\overline{\vec{\Sigma}}\!\cdot\Gamma\!
         (\overline{K}\vec{\tau}N)+(\overline{N}\Gamma\vec{\tau}K)
         \!\cdot\!\vec{\Sigma}\right]
   -f_{\Xi\Sigma K}\left[\overline{\vec{\Sigma}}\!\cdot\!
       \Gamma (\overline{K_{c}}\vec{\tau}\Xi)
     +(\overline{\Xi}\Gamma\vec{\tau}K_{c})\!\cdot\!\vec{\Sigma}\right]
                                         \nonumber\\
 &&-f_{N\!N\eta_{8}}(\overline{N}\Gamma N)\eta_{8}
   -f_{\Lambda\Lambda\eta_{8}}(\overline{\Lambda}\Gamma\Lambda)\eta_{8}
   -f_{\Sigma\Sigma\eta_{8}}(\overline{\vec{\Sigma}}\!\cdot\!\Gamma
       \vec{\Sigma})\eta_{8}
   -f_{\Xi\Xi\eta_{8}}(\overline{\Xi}\Gamma\Xi)\eta_{8}.    \label{Lbar8}
\end{eqnarray}
where the charged-pion mass serves as the mass-scale compensating the effect of the derivative coupling.
The - in total 16 - pseudo-scalar $BB'$-meson vertices are completely fixed by the three nonet coupling constants $(f_8,f_0,\alpha)$
\begin{equation}
  \begin{array}{lll}
   f_{N\!N\pi}               = f_8,                               \ \ \  &
   f_{\Lambda N\!K}          =-\frac{1}{\sqrt{3}}\,f_8(1+2\alpha),\ \ \  &
   f_{N\!N\eta_{8}}          = \frac{1}{\sqrt{3}}\,f_8(4\alpha-1),
      \\
   f_{\Sigma\Sigma\pi}       = 2f_8\alpha,                        \ \ \  &
   f_{\Xi\Lambda K}          = \frac{1}{\sqrt{3}}\,f_8(4\alpha-1),\ \ \  &
   f_{\Lambda\Lambda\eta_{8}}=-\frac{2}{\sqrt{3}}\,f_8(1-\alpha),
      \\
   f_{\Lambda\Sigma\pi}      = \frac{2}{\sqrt{3}}\,f_8(1-\alpha), \ \ \  &
   f_{\Sigma N\!K}           = f_8(1-2\alpha),                    \ \ \  &
   f_{\Sigma\Sigma\eta_{8}}  = \frac{2}{\sqrt{3}}\,f_8(1-\alpha),
      \\
   f_{\Xi\Xi\pi}             =-f_8(1-2\alpha),                    \ \ \  &
   f_{\Xi\Sigma K}           =-f_8,                               \ \ \  &
   f_{\Xi\Xi\eta_{8}}        =-\frac{1}{\sqrt{3}}\,f_8(1+2\alpha).
  \end{array}
                        \label{eq:goct}
\end{equation}
The pseudo-scalar singlet interactions are treated accordingly, leading to the relations
\begin{equation}
   f_{N\!N\eta_{0}}=f_{\Lambda\Lambda \eta_{0}}
    =f_{\Sigma\Sigma\eta_{0}}=f_{\Xi\Xi \eta_{0}}
    =f_{0}.  \label{gsin}
\end{equation}
By the same approach, similar relations are found for the scalar and vector meson Lagrangians and corresponding coupling constants.

\subsection{SU(3) symmetry breaking}\label{ssec:SymBreak}
From the above relations, the advantage of referring to SU(3) symmetry is obvious: for each type of meson (pseudo-scalar, vector, scalar) only four independent parameters are required to characterize their interaction strengths with all possible baryons. These are the singlet
coupling constant $f_0$, the octet coupling constant $f_8\equiv f$, the $F/(F+D)$ ratios, and eventually a mixing angle which relates the physical, dressed isoscalar mesons to their bare octet and singlet counterparts. However, SU(3) symmetry is broken at several levels and SU(3) relations will not be satisfied exactly. An obvious one is the non-degeneracy of the physical baryon and meson masses within the multiplets. As discussed below, this splitting leads to additional complex structure in the set of coupled equations for the scattering amplitudes because the various baryon-baryon ($BB'$) channels open at different threshold energies $\sqrt{s_{BB'}}=m_B+m_{B'}$. At energies $s<s_{BB'}$ a given $BB'$-channel does not contain asymptotic flux but contributes as a virtual state. Thus, $N\Lambda$ scattering, for example, will be modified at any energy by admixtures of $N\Sigma$ channels.

SU(3) symmetry is broken explicitly by the fact that $\Lambda$ and $\Sigma^0$ have the same quark content but are coupled to different total isospin and carry different masses. There is an appreciable electro-weak mixing between the ideal isospin-pure $\Lambda$ and $\Sigma^0$ states~\cite{Dal64}. Exact SU(3) symmetry of strong interactions predicts $f_{\Lambda\Lambda\pi^0}=0$, but $\Lambda$-$\Sigma^0$ mixing results in a weak, but non-zero $\Lambda\Lambda\pi$ coupling constant for the physical $\Lambda$-hyperon, as derived by Dalitz and von Hippel~\cite{Dal64} already in the early days of strangeness physics,
\begin{equation}
  f_{\Lambda\Lambda\pi}=-2\frac{\langle\Sigma^0|\delta M|\Lambda\rangle}
             {M_{\Sigma^0}-M_{\Lambda}}\,f_{\Lambda\Sigma\pi},
\end{equation}
where the $\Sigma\Lambda$ element of the mass matrix is given by
\begin{equation}
   \langle\Sigma^0|\delta M|\Lambda\rangle=
       \left[M_{\Sigma^0}-M_{\Sigma^+}+M_p-M_n\right]/\sqrt{3}.
\end{equation}
Substituting for the physical baryon masses, we find
$ f_{\Lambda\Lambda\pi}=c_b\,f_{\Lambda\Sigma\pi}$ where $c_b=-0.0283...$ is the symmetry breaking coefficient. From the nucleon-nucleon-pion part of the interaction Lagrangian
(\ref{Lbar8}), we find
\begin{equation}
    (\overline{N}\vec{\tau}N)\!\cdot\!\vec{\pi} =
    \overline{p}p\pi^0-\overline{n}n\pi^0+\sqrt{2}\,\overline{p}n\pi^+
                      +\sqrt{2}\,\overline{n}p\pi^-,
\end{equation}
and the neutral pion is seen to couple with opposite sign to neutrons and protons. This implies that the non-zero $f_{\Lambda\Lambda\pi^0}$ coupling produces considerable deviations from charge symmetry in $\Lambda p$ and $\Lambda n$ interactions. Obviously, $\Lambda$-$\Sigma^0$ mixing also gives rise to non-zero $\Lambda\Lambda$ coupling constants for any (neutral) isovector meson, in particular for the $\rho$ vector meson and the $\delta/a_0(980)$, the isovector member of the $0^+$ meson octet, respectively but the latter couplings give rise to much smaller effects.

As seen from Eq.(\ref{eq:goct}), $\Lambda$ and $\Sigma$ hyperons are coupled by strong interactions leading to a strangeness- and isospin/charge-conserving dynamical mixing of $N\Lambda$ and $N\Sigma$ channels. Again, corresponding effects are present also for vector-isovector and scalar-isovector mesons. As a result, channels of the same total strangeness $S$ and fixed total charge $Q$ are coupled already on the level of the elementary vertices in free space. In asymmetric nuclear matter, this effect is enhanced through the isovector mean-field produced by the $\rho_0$ and the $\delta_0$ meson fields. This leads to a density dependent $\Lambda\Sigma^0$ mixing being proportional to the isovector mean-field $U_1\sim \rho_1$ and determined by the isovector baryon density $\rho_1=\rho_n-\rho_p$ plus possible contributions from charged hyperons. As a result, $\Lambda$ and $\Sigma^0$ become quasi-particles with mass eigenstates different from the flavor eigenstates. The total isospin is of course conserved by virtue of the background medium.

\subsection{Interactions and self-energies}
On the tree-level the octet interaction Lagrangians discussed in the previous section correspond to standard One Boson Exchange amplitudes. The full interaction amplitudes require the solution of the Bethe-Salpeter equation. As in \cite{Dejong:1998}, we reduce the problem to an effective Lippmann-Schwinger equation by projecting out the time-like components by the Thompson method. However, different from the better known SU(2) nucleon-nucleon case \cite{Dejong:1998,Machleidt:1989}, the full SU(3) problem leads to a set of coupled equations describing interactions within the set of channels of fixed strangeness $S=0,-1\cdots -4$ and total charge $Q$, respectively, where the latter corresponds to fix the third component of the total isospin. Hence, in momentum representation and after performing a partial wave decomposition we have to solve the coupled integral equation
\be\label{eq:Tmat}
T_{ab}(q,q')=\tilde{V}_{ab}(q,q')+\sum_c{\frac{1}{2\pi^2}\int{dkk^2V_{ac}(q,k)Q^{(c)}_F(k,s)G_c(k,s)T_{cb}(k,q') } }
\ee
where the indices $a,b,c$ account for all relevant quantum numbers in the corresponding flavor and partial wave channels. $G_c$ denotes the Green's function in the intermediate channel $c$, to be evaluated at the on-shell total center-of-mass energy $s$. In nuclear matter, the projector $Q^{(c)}_F=\Theta(k^2_1-k^2_{F1})\Theta(k^2_2-k^2_{F2})$ accounts by step-functions for the Pauli-principle by blocking the occupied states inside the Fermi spheres $k_{F1,2}$ of particle 1 and particle 2. In practice, we use angle-averaged Pauli-projectors. In free space we have $Q^{(c)}_F\equiv 1$. Additional in-medium modifications are introduced by self-energy, mainly modifying the propagators.

The Born-term includes also the kinematical thresholds at $s=s_a,s_b$
\be
\tilde{V}(q,q')=V(q,q')\Theta(s-s_{a})\Theta(s-s_{b})
\ee
thus accounting explicitly for SU(3) symmetry breaking. By a properly chosen quadrature method, Eq.(\ref{eq:Tmat}) is conveniently transformed into a numerically easy to solve linear system, see e.g. \cite{Machleidt:1989}. The bare coupling constants $\left\{f_0,f_8,\alpha\right\}_{p,v,s}$ must be determined phenomenologically by fit to data where at present the scarce data base inhibits a precise determination. A brief discussion of the approach is found elsewhere in these proceedings \cite{MD:2015} where also our results for free space cross sections are shown.

\begin{figure}[tbh]
\begin{center}
\includegraphics[width=10cm]{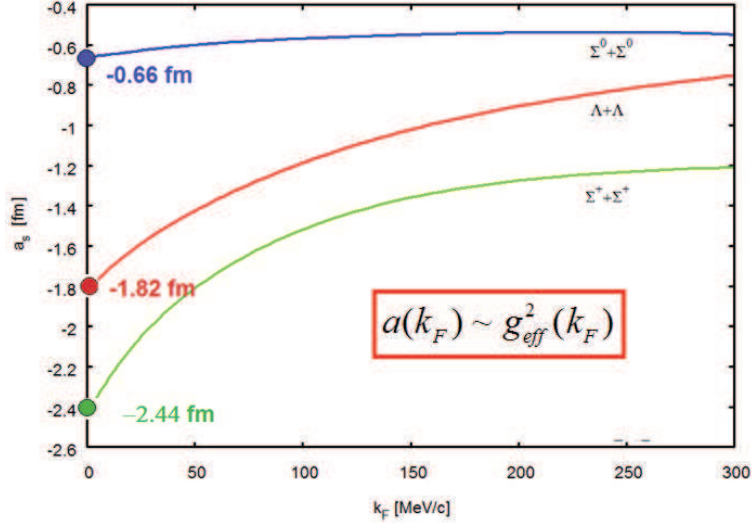}
\caption{In-medium Singlet-Even ($SE$) s-wave scattering lengths for the $S=-2$ double-hyperon channels as a function of the Fermi-momentum $k_F\sim \rho^{1/3}$ (see also \protect\cite{MD:2015}). Our free-space values are indicated on the y-axis.}
\label{fig:Fig3}
\end{center}
\end{figure}

The low-energy parameters are convenient measures to characterize the interactions strength. As an example, we consider the effective range expansion in the $S=-2$ singlet-even channels which are of particular interest for $S=-2$ hypernuclei. Thus, for $q\to 0$,
\be
\frac{1}{q}\tan{\delta_{SE}(q,k_F)}\sim -a_{SE}(k_F)\left(1+\frac{1}{2}q^2 a_{SE}(k_F)r_{SE}(k_F)+\cdots\right)
\ee
where $r_{SE}$ denotes the effective range parameter. The scattering length is related to the volume integral of the K-matrix,  $a_{SE}(k_F)=\frac{2\mu}{\hbar^2 4 \pi}K_{SE}(0,k_F)_{|\ell=0}$, where $\mu$ is the reduced mass and $T=(1-iK)^{-1}K$. Corresponding relations are found also for the other interaction channels. Thus, up to a normalization factor the scattering lengths correspond to in-medium effective coupling constants. As seen in Fig.(\ref{fig:Fig3}) the interaction strength decreases rapidly in all channels with increasing density approaching a plateau at densities close to nuclear equilibrium. Overall, the scattering lengths are only about 10\% of the values found in the $NN$-system, for the latter see e.g. \cite{Machleidt:1989}, reflecting the much weaker $YY$ interaction.

With the T- and K-matrices available, we have access to the full scattering amplitude. For nuclear structure work we are interested in the first place on the G-matrix as introduced by Brueckner which is nothing than the in-medium K-matrix. While the natural representation of the K~-matrix is the \textit{singlet/triplet-even/odd} formalism, for nuclear matter calculations the explicit representation in terms of spin-isospin operators is more convenient. The two representation are related by an orthogonal transformation among the two operator sets. Thus, for each isospin doublet introduced in Eq.(\ref{eq:doublets}) we obtain in non-relativistic reduction a structure familiar from the nucleon sector,
\bea
R^{BB'}(\xi,k_F)&=&
\sum_{S,I=0,1}{R^{BB'}_{SI}(\xi,k_F)\left(\vec{\sigma}_B\cdot\vec{\sigma}_{B'}\right)^S\left(\vec{\tau}_B\cdot\vec{\tau}_{B'}\right)^I  } \nonumber \\ &+&\sum_{I=0,1}{\left(R^{BB'}_{TI}(\xi,k_F)\mathbf{S_{12}}+R^{BB'}_{LI}(\xi,k_F)\mathbf{L\cdot S}\right)\left(\vec{\tau}_B\cdot\vec{\tau}_{B'}\right)^I }
\eea
and also spin-orbit and rank-2 tensor terms are displayed. The amplitudes $R_{SI}(\xi,k_F)$ are given either in momentum space ($\xi={\mathbf{k_1,k_2}}$) or, if a static potential picture is an acceptable approximation, in coordinate space ($\xi={\mathbf{r_1,r_2}}$). For a given $(BB')$-combination some of the $R_{SI}$ may vanish. Formally, also the interactions involving the $\Sigma$~-isotriplet can be cast into such a form, except that one (or both) Pauli-type isospin operators have to be replaced by the proper spin-1 operator,  $\vec{\tau}_B\to \vec{T}_\Sigma$. In general, the $R_{SI}$ are functionals of the field operators, as studied in detail in our field-theoretical DDRH approach \cite{Fuchs:1995,Lenske:2004,Dejong:1998,Fedo:2014} and exploited later by many other groups by means of phenomenological mean-field parameterizations, e.g. \cite{lala}.  In a single-hyperon spin-saturated nucleus the hyperon mean-fields are then given by
\be
U^{YA}=U^{YA}_{0}+U^{YA}_{L0}\vec{\ell_Y}\cdot\vec{\sigma_Y}+\left(U^{YA}_{1}+U^{YA}_{L1}\vec{\ell_Y}\cdot\vec{\sigma_Y}\right) \tau^0_Y
\ee
where the central isoscalar and isovector potentials $U^{YA}_{0,1}(\rho_{0,1})$ are determined by the isoscalar and isovector nuclear densities, $\rho_{0,1}=\rho_n\pm\rho_p$, respectively. Similar expressions hold also for the spin-orbit terms. An open problem for the future is the consistent treatment of three- and many-body forces, urgently needed for neutron star physics, see e.g. \cite{Yama:2014}.

\begin{figure}[tbh]
\begin{center}
\includegraphics[width=12cm]{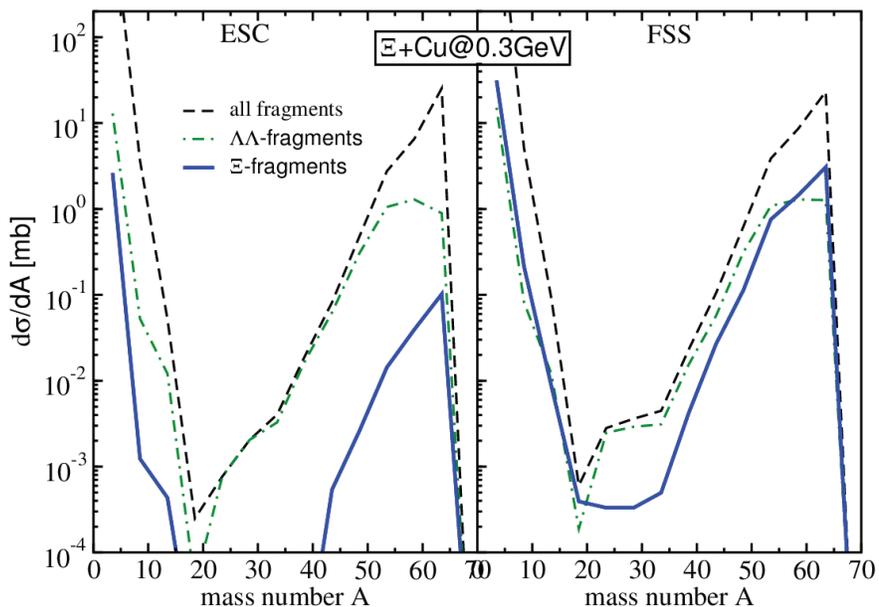}
\caption{Mass distributions of all fragments (dashed curves), double-$\Lambda$ fragments (dot-dashed curves)and $\Xi$ hyperfragments (thick solid curves). Two sets of calculations are shown, namely one using the ESC (left panel) and the FSS model (right panel). The yields are from inclusive $\Xi$-induced reactions at a beam energy $E_\Xi=0.3$~GeV corresponding to the momentum $p_\Xi=0.937$~GeV/c (see also \protect\cite{Gaitanos:2014}).}
\label{fig:Fig4}
\end{center}
\end{figure}

\section{Strangeness production in heavy ion collisions and antiproton annihilation}\label{sec:HIC}

Heavy ion collisions at relativistic energies are an interesting and relatively new path to strangeness and hypernuclear production by strong interaction. Experimentally, the HypHI group \cite{HypHI} at GSI has initiated a strong activity in that direction, showing the feasibility of hyperon and hypernuclear production by heavy ions. On the theory side, major activities in this direction are being made by the Frankfurt-Mainz collaboration \cite{Botvina:2013,Botvina:2014} and our own investigations. In a series of paper \cite{Gaitanos:2009,Gaitanos:2011,Larionov:2012,Gaitanos:2012,Gaitanos:2014} we have been studying hyperon and hypernuclear production in proton and antiproton-induced reaction and in heavy ion collisions leading to fragmentation into light hypernuclei. In \cite{Gaitanos:2014} we have for the first time investigated the role and possible observable effects of different YN-interaction
models on the strangeness dynamics of hadron-induced reactions. The main focus of that work was to study to what extent differences in the elementary interaction model are reflected in hypernuclear observables. As appropriate cases, being also of experimental interest for \panda, we have chosen antiproton annihilation leading to hyperon production and the interactions of secondary beam of $\Xi$ hyperons with nuclei, thus studying specifically the $S=-2$ sector of the theory. Improved parameterizations for the cross sections of the S=-2 YN-channels are based on the microscopic calculations of the Nijmegen group by Rijken et al.~\cite{Nijmegen} and by Fujiwara et al.~\cite{Niigata} and implemented into the Giessen-BUU ($GiBUU$) transport model code~\cite{Buss:2011mx}. In brief, our results are showing strong dynamical response on the underlying $\Xi N$-interactions. Differences are clearly visible in the yields of multi-strangeness hypernuclear production in low-energy $\Xi$-induced reactions. Hence, the proposed production experiments may lead to strong constraints on the high strangeness YN and YY interactions.

Antiproton-induced primary reactions and the subsequent reactions with a secondary $\Xi$-beam are described by the well-tested relativistic Boltzmann-Uheling-Uhlenbeck (BUU) transport approach. The kinetic equations are numerically realized within the GiBUU transport model~\cite{Buss:2011mx}. Thus, we solve the transport equation
\begin{equation}
\left[
k^{*\mu} \partial_{\mu}^{x} + \left( k^{*}_{\nu} F^{\mu\nu}
+ m^{*} \partial_{x}^{\mu} m^{*}  \right)
\partial_{\mu}^{k^{*}}
\right] f(x,k^{*}) = {\cal I}_{coll}
\quad .
\label{rbuu}
\end{equation}
which describes the dynamical evolution of the one-body phase-space
distribution function $f(x,k^{*})$ for the hadrons under the influence of a
hadronic mean-field (l.h.s. of Eq.~(\ref{rbuu})) and binary collisions (r.h.s. of Eq.~(\ref{rbuu})). The mean-field enters via the in-medium self-energies,
$\Sigma^{\mu} = g_{\omega}\omega^{\mu} + \tau_{3}g_{\rho}\rho_{3}^{\mu}$ and
$\Sigma_{s} = g_{\sigma}\sigma$, into the transport equation, contained in
the kinetic $4$-momenta $k^{*\mu}=k^{\mu}-\Sigma^{\mu}$ and effective (Dirac)
masses $m^{*}=M-\Sigma_{s}$. The various meson fields are obtained from the standard Lagrangian equations of motion~\cite{qhd}, extended to non-linear self-interactions of the $\sigma$ field ~\cite{lala}. A schematic approach is used for the meson-hyperon mean-field couplings which were obtained from the nucleonic sector by a simple quark-counting rules. Of special interest, however, is the collision term. Besides including all necessary binary processes for (anti)baryon-(anti)baryon, meson-baryon and meson-meson scattering and annihilation~\cite{Buss:2011mx}, the most important ingredients are the new $\Xi$N cross section parameterizations.

The main results of our investigation in \cite{Gaitanos:2014} are found in Fig.(\ref{fig:Fig4}). The most remarkable feature is the
strong dependence of the distributions of $\Xi$-bound hypernuclei on the
underlying $\Xi$N model, although both interactions are leading to almost identical $S=-1$ distributions \cite{Gaitanos:2014}. While for the ESC model the production
probability of bound $\Xi$-matter is relatively low, the production of multi-strange $\Xi$-systems considerably is considerably enhanced when using the FSS
interactions. The differences are of such a magnitude that a sizeable spread is predicted for the $S=-2$ production cross sections. It would be therefore a challenge to measure exotic
multi-strangeness hypermatter at \panda in the future, in order to better
constrain the experimentally still unknown and theoretically obviously very controversial interactions in the $S=-2$ channel, eventually ruling out certain approaches to YN and YY interactions.

\begin{figure}[tbh]
\begin{center}
\includegraphics[width=14cm]{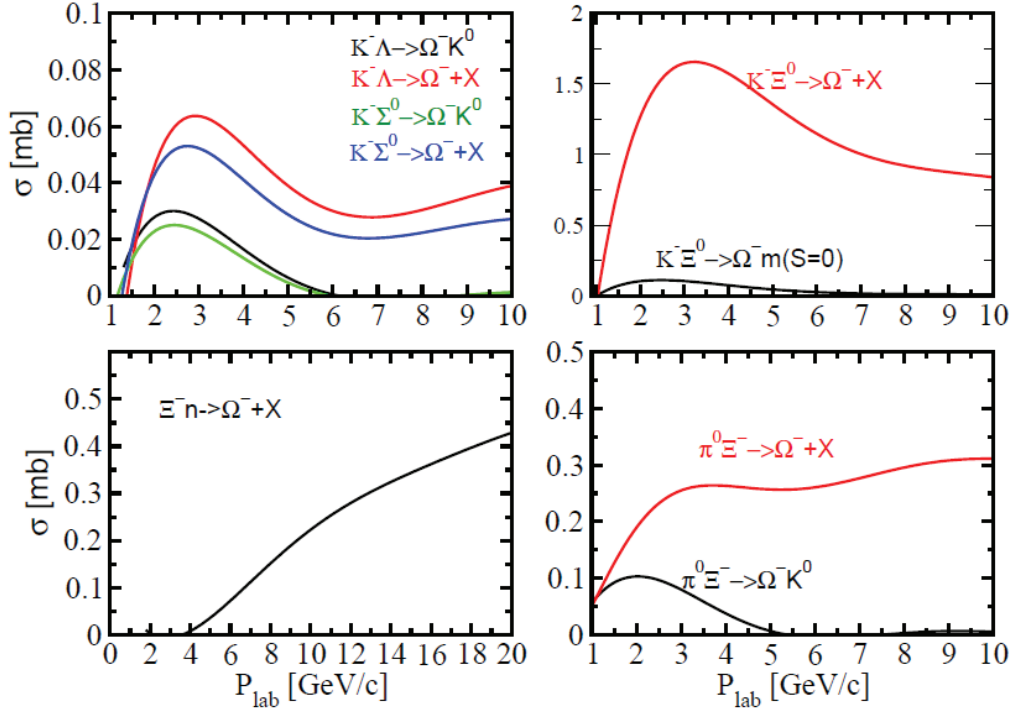}
\caption{In-medium $\Omega^-$ production by sequential strangeness accumulation reactions. The momentum of the incoming baryon, produced as a secondary beam, is denoted by $p_{Lab}$.}
\label{fig:Fig5}
\end{center}
\end{figure}

It is tempting to extend the investigations further into the sector of higher strangeness, both theoretically and experimentally. Possible routes are to study the production of nuclear multi-strangeness states of octet hyperons, e.g. possible $S=-3,-4$ configurations built by combined $\Lambda\Xi$ and double-$\Xi$ states, respectively. Also the question of the possible existence of bound $\Sigma$-nuclei is still undecided. An even more interesting direction is to leave the octet sector and extend the approach into the $3/2+$-baryon decuplet, containing on the one hand excited states of the octet baryons but also a completely new particle, the $\Omega^-$ hyperon. The $\Omega^-$ is made out of three s-quarks, carrying a total strangeness of $S=-3$. Thus, it is the baryon with largest possible strangeness content. As an outlook we report here on first steps towards $\Omega^-$ production in $\bar p A$ annihilation reactions. Already the description of the production of a  $\Omega^-$ baryon is a challenge to theory. Direct
$p\bar p\to\Omega \bar{\Omega}$ production is OZI-suppressed because the creation of three $s\bar s$ pairs out of the vacuum would be required. Hence, sequential processes with step-wise strangeness accumulation are more likely. First results for such in-medium cross sections are shown in Fig.(\ref{fig:Fig5}). Inserting these cross sections into a full-scale transport calculation will allow to explore high-strangeness production processes quantitatively. Such calculations are in preparation.

\section{Summary}\label{sec:sum}
Strangeness physics is a field of particular interest for our understanding of baryon dynamics in the very general context of low-energy flavor physics. The rich spectrum of phenomena is extending from strangeness production on elementary reactions on the nucleon, probing the full variety of probes, ranging from virtual or real photons and leptons to elementary hadronic reactions and, as new challenges, antiproton-nucleus and nucleus-nucleus reactions. In this contribution, the richness of phenomena was illustrated on selected examples. An outlook to extensions of hypernuclear physics into the region of high strangeness $|S|>2$ was given.
\vspace{5mm}
\paragraph{Acknowledgement:}Supported by DFG, contract Le439/9, BMBF, contract 05P12RGFTE, GSI Darmstadt, and Helmholtz International Center for FAIR.

\end{document}